\documentclass{elsart}

\usepackage{amssymb}
\usepackage{graphicx}
\usepackage{epstopdf}

\begin{document}

\begin{frontmatter}



\title{Particle production and nonlinear diffusion 
in relativistic systems}


\author{Georg Wolschin}
\footnote{Email: wolschin@uni-hd.de}
\address{Institut f\"ur Theoretische Physik der Universit\"at 
Heidelberg, Philosophenweg 16, D-69120 Heidelberg, Germany}

\begin{abstract}
The short parton production phase in high-energy heavy-ion
collisions is treated analytically  as a nonlinear diffusion process. 
The initial buildup of the rapidity density 
distributions of produced charged hadrons 
within $\tau_{p}\simeq $ 0.25 fm/c occurs in three sources
during the colored partonic phase. 
In a two-step
approach, the subsequent
diffusion in pseudorapidity space during the interaction
time of $\tau_{int} \simeq$ 7-10 fm/c (mean duration of the collision)
is essentially linear as expressed
in the Relativistic Diffusion Model (RDM) which yields
excellent agreement with
the data at RHIC energies, 
and allows for predictions at LHC energies.
Results for d+Au are discussed in detail.

\end{abstract}

\begin{keyword}
Relativistic heavy-ion collisions  \sep Particle production \sep Nonlinear diffusion model
\sep Pseudorapidity distributions

 \PACS 25.75.-q \sep 24.60.Ky \sep 24.10.Jv 
\end{keyword}
\end{frontmatter}

\section{Introduction}
\label{intro}
A time-dependent analytical description of particle production
from the available relativistic energy in heavy-ion collisions
is of considerable interest. In particular,
the accurate modeling of transverse momentum and rapidity 
distribution functions for produced
particles is a basic requirement
in attempts to understand the relevant partonic and hadronic physical
processes. Analytically solvable models offer 
transparent approaches to the problem, including the possibility
to extrapolate to other incident energies, such as
from RHIC energies
$\sqrt{s_{NN}}$=19.6 -- 200 GeV to LHC, $\sqrt{s_{NN}}$= 5.52 TeV.
 
In this work I propose a nonequilibrium-statistical (and hence,
time-dependent) approach that provides
analytical solutions for pseudorapidity distributions of produced particles.
It is based on a nonlinear diffusion equation in 
rapidity space, which accounts for the explicit dependence of
the diffusion coefficient on the rapidity density in the initial
short ($\tau_{p} \simeq $ 0.25 fm/c) partonic phase of the collision 
when most of the particles are produced. This is followed
by the somewhat more extended phase of color neutralization \cite{kns07},
and a long-lasting color-neutral pre-hadronic or hadronic phase
with rapid expansion of the system.

Three sources for particle production
are considered, two for initial rapidities 
close to the beam values, and a third central source that arises
mostly from gluon-gluon collisions.
I consider the transition from the initial, highly nonlinear
partonic phase to the subsequent, essentially linear phase. 
Other investigations
such as \cite{kah05}, and references to numerical
simulations therein, corroborate the short duration of the parton-production
phase, and the long duration of the subsequent recombination
(pre-hadronic) and hadronic phase. According to parton-cascade models
such as \cite{elg96}, the pre-hadrons
are color singlets that are generated from quark and gluon
recombination in a statistical coalescence process. They decay 
into the final hadrons
according to their relative phase-space weights. 
Many numerical approaches to the problem use string models in the initial phase,
and final-state hadronic collisions, but often hadronic rescattering is not considered.

The second phase lasts about
$\tau_{int} \simeq$ 7-10 fm/c (depending on the system,
the incident energy, and the centrality). Phenomenologically, it
is well accounted for in a Relativistic Diffusion Model (RDM)
based on a linear Fokker-Planck equation
with constant diffusion coefficient, plus fast collective expansion.
The RDM had been developed some time ago \cite{wol99,biy02,wols06,wols07,kw07}
and compared in detail with data on net proton rapidity distributions at
SIS, AGS, SPS, and RHIC energies,
and with produced-particle distributions. For pseudorapidity distributions
of produced charged hadrons, a
$\chi^{2}$-optimization of the Jacobi-transformed analytical solutions
yields very precise agreement with the available RHIC data provided the
midrapidity source for particle production is taken into account.

For net-proton rapidity distributions at RHIC energies, there have been
several investigations of nonlinear effects within the
diffusion approach. 
A nonlinearity in the drift coefficient that secures the
correct Maxwell-Boltzmann equilibrium limit for $t \rightarrow \infty$
has been investigated in \cite{alb00},
but the deviations from linearity are small.

The strong effect of a nonlinearity in the
diffusion coefficient that also persists over the full interaction time
of typically 10 fm/c \cite{lis05,st99}
as a consequence of the introduction of 
non-extensive statistics \cite{tsa88} had been investigated
in \cite{alb00,wol03} for net protons in heavy-ion collisions, and in \cite{ryb03}
for produced particles in $p\bar{p}$-collisions. It seems that this is a way to
account for the collective expansion of the system without
considering it explicitly. If one includes an explicit treatment of collective
expansion \cite{wol06}, however, there is no need to introduce
non-extensive statistics when comparing the RDM-results
with data, the linear evolution after the parton-production
phase yields excellent results for both net baryons, and produced particles.

The linear diffusion model had also been proposed in order to calculate and
predict transverse energy distributions of hadrons \cite{wol96}, and more
recently to calculate transverse momentum distributions of identified
hadrons (neutral pions, as well as negative pions and kaons)
in a nonequilibrium-statistical approach  
including radial flow \cite{suz07}.

Equilibrium-statistical models \cite{bm01,an06,bec01} account in remarkable 
detail for relative production rates 
of produced particles at central rapidity with only the
temperature and the chemical potential as parameters. 
Due to the lack
of time dependence - and consequently,
of nonequilibrium-statistical effects -
a relevant ingredient is, however, missing if one
aims at the precise modeling of distribution functions.

The gradual approach of the system towards statistical equilibrium
in the course of a relativistic heavy-ion collisions
is presented in a schematic analytical model in this work.
I start with the nonlinear diffusion model
as expected to be valid during the parton-production phase in 
Section \ref{nonlde}, followed by a consideration of the so-called
source solution of the nonlinear problem in Section \ref{source}.
For sufficiently large times, the initial power-law behaviour
during the nonlinear parton-production phase is superseded by the
essentially linear diffusion (RDM-) phase in rapidity
space which produces gaussian tails in pseudorapidity
space, Section \ref{lin}. 

The late-stage time evolution is discussed in a 
comparison with RHIC data for the asymmetric d + Au- system,
which indeed show the gaussian tails.
In an asymmetric system like d+Au,
the nonequilibrium effects are visible more directly
than in case of symmetric systems such as Au + Au. The
interaction ceases long before statistical equilibrium
with respect to the variable rapidity 
is reached. The conclusions are drawn in Section \ref{conc}.

\section{The nonlinear diffusion equation}
\label{nonlde}
The origin of diffusion during and after particle production 
in a heavy-ion reaction at relativistic energies is found
in momentum space, through random momentum kicks of 
the produced particles -- partons  in the soft-gluon field in the early stage,
prehadrons and hadrons in later stages.
Diffusion in coordinate space appears as a secondary effect. 

The corresponding fluctuations can be seen in rapidity and
pseudorapidity distributions of produced particles 
\cite{wol99,biy02,wols06,wols07,kw07}, as well 
as in transverse energy and momentum distributions \cite{wol96,suz07}.
To provide an analytical treatment of the problem in both
early and late stages, I confine
the present work to rapidity space, with the lorentz-invariant
rapidity  $y=0.5\cdot \ln((E+p)/(E-p))$, and a subsequent Jacobian
transformation to pseudorapidity $\eta$ that is required
for the comparison to the available data for produced charged hadrons.

To incorporate the early parton-production phase into the
relativistic diffusion model  \cite{wol99,biy02,wols06},
a dependence of the diffusion coefficient on the initially
very high rapidity density $R(y,t)$ has to be considered,
$D_{y}\rightarrow D_{y}(R)$, such that the linear
transport equation in rapidity space that I investigated in 
\cite{wol99} is replaced by 
\begin{equation}
\frac{\partial}{\partial t}R(y,t)=-\nabla_{y}\Bigl[J(y)R(y,t)\Bigr]+
\nabla_{y}D_{y}(R(y,t))\nabla_{y}R(y,t).
\label{nlfpe}
\end{equation}
 The drift
term $J(y)$ governs the gradual approach of the
mean values towards statistical equilibrium.
The diffusion coefficient $D_{y}(R)$ depends on the rapidity
density and hence, the equation is generally highly
nonlinear. It is therefore expected to account not only for the
long-lasting, essentially linear diffusive phase as in \cite{wol99,biy02,wols06},
but also for the partonic 
initial phase of high rapidity density. In this short-lived phase the major
part of particle production with rapidly rising norm
of the distribution function takes place. The rising norm
is phenomenologically accounted for in this work
by letting the integration constant in
Eq.(\ref{nlfpe}) depend on particle number.

In case of the linear RDM \cite{wol99,wol96} with $D_{y}(R)=D_{y}$=const., 
I had assumed an instant production of the
particles in the three sources, and subsequent diffusion in $y-$space during
the interaction time. This initial condition is exactly fulfilled only
for net baryons \cite{wol99}. However, for produced charged
hadrons, it also yields extremely precise results when compared
\cite{biy02,wols06,kw07} in detail to the data. An explicit treatment of the
short nonlinear parton production phase with a strong dependence of the
diffusion coefficient on the distribution function should therefore
preserve the model features of the subsequent, essentially 
linear diffusive phase.

To account for the strong correlation between diffusion
coefficient and rapidity density distribution in
the initial high-density particle production phase,
I propose a dependence on a power $\kappa$
of the rapidity density according to
\begin{equation}
D_{y}[R(y,t)]=D_{y}^{p}\cdot R(y,t)^{\kappa}
\label{nld}
\end{equation}
with $D_{y}^{p}$ the rapidity diffusion constant
in the particle production phase.
For certain critical exponents $\kappa$, analytical solutions of the
diffusive part of the transport equation can be obtained.

In a two-step approach, the subsequent -- probably mostly pre-hadronic
and hadronic --
evolution in pseudorapidity space during the interaction
time of $\tau_{int} \simeq$ 7-10 fm/c (mean duration of the collision;
see \cite{lis05} )
is accounted for
in the Relativistic Diffusion Model  [1] (RDM, $\kappa=0$) 
with a linear drift term
$J(y)=(y_{eq}- y)/\tau_{y}$ governed by the rapidity relaxation time
$\tau_{y}$ and the equilibrium value of the rapidity $y_{eq}$. The
diffusion term is here $\propto  
D_{y}\frac{\partial^2}{\partial y^2}R(y,t)$. 

The diffusion constant $D_{y}$ in this 
phase is significantly smaller than
the value of $D_{y}^{p}$ in the short production phase with
a large number of (partonic) degrees of freedom. The linear model
with instant particle production yields
excellent agreement with d+Au, Cu+Cu and Au+Au data at RHIC energies,
including the detailed centrality dependence \cite{wols06}.\\

The initial short, highly nonlinear phase of parton production
occurs within $\tau_{p}\simeq $ 0.25 fm/c \cite{kah05} in two
beam-like sources, and a central source in rapidity space.
Since the time scale for particle
production in all three sources is faster than 
the one for the nonlinear diffusion, it turns out that the
particle content in the power-law tails remains small: 
there is little spread of the distribution 
function in rapidity space in this initial phase.

The mathematical treatment of the initial nonlinear
phase is confined to the central source 
as an example. The beam-like sources are then
dealt with in an analogous way, but with the freedom to choose different
diffusion coefficients because 
the production mechanisms in the valence-quark
dominated beam-like regions of rapidity space are different
from the central region with few valence quarks at RHIC or LHC energies.
Kinematic constraints in the beam-like sources expected at high
absolute values of rapidity are not considered here.

During the short production phase, the drift in $y-$space is not yet
pronounced, and  I therefore treat here only the diffusive part of
the nonlinear transport equation, $R(y,t) \rightarrow P(y,t)$ with
\begin{equation}
\frac{\partial}{\partial t}P(y,t)=D_{y}^{p}
\nabla_{y}P(y,t)^{\kappa}\nabla_{y}P(y,t).
\label{nlfped}
\end{equation}
The solution of this equation at the end of the nonlinear production
phase ($t=\tau_{p}$) can then be used as initial condition for the subsequent
linear diffusive time evolution treated in the 
Relativistic Diffusion Model (RDM)
\cite{wol99,biy02,wols06,wols07,kw07},
\begin{equation}
R_{0}(y,0)=P(y,\tau_{p}) .
\label{ini}
\end{equation}
With $t^{*} = t \cdot D_{y}^{p}$ the nonlinear diffusion equation becomes
\begin{equation}
\frac{\partial}{\partial t^{*}}P(y,t^{*})=
\nabla_{y}P(y,t^{*})^{\kappa}\nabla_{y}P(y,t^{*}).
\label{nlp}
\end{equation}
This equation has been 
extensively studied in many diverse areas
of science, and a large amount of mathematical literature exists, \cite{pat59,phi60,
tuc76,per77,lac82,kat84,hil89,gan98,cha99},
and references therein. It has mostly been considered for positive values of $\kappa$
such as $\kappa=1$ for thin saturated regions in porous media, $\kappa \geq 1$ 
for the percolation of gas through porous media, $\kappa=3$ for thin films spreading under
gravity, and $\kappa=6$ for radiative heat transfer by Marshak waves.
Many problems are dealt with in only one (spatial) dimension, analogous to
the present work which is confined to one (momentum-like) dimension.
It has also been established that the number of exact solutions
is limited.  Solutions for several power-law diffusivities with negative $\kappa$
are known \cite{hil89}, in particular, for $\kappa = -1/2, -1, -4/3, -3/2$ and $ -2$.

The physically most interesting behaviour occurs in the region of small
diffusion coefficients D(R), where a moving boundary may exist.
The behaviour of solutions for positive
and negative values of $\kappa$ is distinctively different. Depending
on the specific values for the constants of integration (see below), 
a free boundary may occur for $\kappa$ = 1 that has a finite 
gradient and moves with finite
velocity, similarly for other positive values of $\kappa$, but for $\kappa > 1$,
the gradient at the boundary becomes infinite.

For $\kappa = -1$
and some other negative values,
however, there is an instantaneous spread without a free
boundary, as was treated by Pattle \cite{pat59}, Pert  \cite{per77}  
Hill \cite{hil89} and
others for instantaneous heat deposition in a medium
with concentration-dependent diffusion coefficient.
This situation is in some respect analogous to
the initial parton production from the
available energy in a relativistic heavy-ion collision. Here the
initial rapidity density distribution of the created partons
should not have a free boundary in
rapidity space, except for kinematical constraints.

Hence I investigate solutions for negative values
of $\kappa$ emphasizing $\kappa = -1$. An exact solution
is not only useful to test the accuracy of numerical results, but 
it is also important to understand and describe the physical
behaviour of the system. In particular, the analytical solution at the end of
the initial nonlinear phase may then be used as initial condition for the
subsequent, essentially linear diffusion process in $y-$space which
can be modeled analytically within the given simple but
successful RDM-framework.

The majority of known exact solutions of the nonlinear problem
are so-called similarity solutions \cite{pat59,phi60,tuc76,per77,lac82,kat84,hil89,gan98}:
With an assumed functional form of the solution, the partial differential
equation reduces to an ordinary differential equation, or to a partial
differential equation of lower order, which can then be integrated
in closed form under certain conditions. The similarity solution of (\ref{nlp})
for $t^{*} \rightarrow t$  is written as
\begin{equation}
P(y,t)=y^{{2\lambda}/{\kappa(1+\lambda)}}\Phi(\xi)
\label{sim}
\end{equation}
with
\begin{equation}
\xi=y^{1/(1+\lambda)}/t^{1/2}
\label{xi}
\end{equation}
and $\lambda$ is an arbitrary constant with $\lambda \ne -1$. Inserting this
ansatz into the nonlinear differential equation (\ref{nlp}) yields first integrals
for two values of $\lambda$ (for $\kappa = -1$ and $-2$, only one value of $\lambda$)
\begin{equation}
\lambda_{1}=-\kappa/(\kappa+2) 
\label{l1}
\end{equation}
\begin{equation}
\lambda_{2}=-\kappa/(\kappa+1) 
\label{l2}
\end{equation}
which have been given by Hill and Hill (1990) \cite{hil89} as
\begin{equation}
\frac{{\Phi^{\kappa}\Phi^{'}}}{\xi}-\frac{2\Phi^{\kappa+1}}{(\kappa+2)\xi^2}+
\frac{2\Phi}{(\kappa+2)^2}=C_{1}
\label{ph1}
\end{equation}
\begin{equation}
\frac{{\Phi^{\kappa}\Phi^{'}}}{\xi}-\frac{(2\kappa+3)\Phi^{\kappa+1}}{(\kappa+1)^{2}\xi^2}+
\frac{\Phi}{(2\kappa+1)^{2}}=C_{1}
\label{ph2}
\end{equation}
with $\Phi^{'}=\partial \Phi/\partial \xi$. For vanishing constants of integration 
$C_1=0$, the so-called "source" and "dipole" solutions arise from these first
integrals. For finite $C_{1}\not= 0$, a number of exact solutions for special
values of $\kappa$ have also been derived \cite{hil89}.
Here I investigate the source solution for 
$C_{1}=0$
\begin{equation}
\Phi=(C_{2}\xi^{2\kappa/(\kappa+2)}-\frac{\kappa \xi^{2}}{2(\kappa+2)})^{1/\kappa}.	
\label{solph}
\end{equation}
This solution
is not defined for $\kappa=-2$, and becomes singular for $\kappa < -2$.
Hence, a solution without a free boundary as required for the initial
rapidity diffusion problem can in principle only occur for
$-2<\kappa<0$. The desired solution for $\kappa=-1$ 
(and hence, $\lambda=\lambda_{1}=1$)
obeys the first-order partial differential equation
\begin{equation}
\frac{\Phi^{'}}{\xi\Phi}-\frac{2}{\xi^2}+
{2\Phi}=0
\label{s1}
\end{equation}
and the solution of the nonlinear diffusion equation 
\begin{equation}
P(y,t)=\Phi(\xi)/y          
\label{sp}
\end{equation}
with $\xi=(y/t)^{1/2}$ becomes
\begin{equation}
P(y,t)=[C_{2}y\xi^{-1}+y\xi^{2}/2]^{-1}
\label{snl}
\end{equation}
where $C_{2}$ is the constant of integration.
\section{The source solution in particle production}
\label{source}
The solution of the nonlinear
diffusion problem in a high-density phase
such as during parton production in rapidity space has thus been reduced to
\begin{equation}
P(y,t)=[C_{2}t+y^{2}/(2t)]^{-1}. 
\label{snld}
\end{equation}
To account for the increasing norm of the
rapidity distribution function during the rapid parton production process,
I let the integration constant $C_{2}$ depend on the particle number.
Since the particle number is likely to increase exponentially with
time during the first collision phase where $t\equiv \hat t<\tau_{p}\simeq 0.25$ fm/c,
I  take the particle-number dependence into account phenomenologically 
by choosing the integration constant in the denominator as
\begin{equation}
C_{2}=\exp(-\hat t ).
\label{c2}
\end{equation}
In the present two-step model, all particles are assumed to be
produced until $t=\tau_{p}$ where the nonlinear production phase
turns into an (essentially linear) diffusion process with $\kappa=0$.
That is, for
$C_{2}=\exp(-\tau_{p})$ the norm of the solution reaches 
its maximum value
\begin{equation}
\int P(y,t=\tau_{p}) dy=1.
\label{nsnld}
\end{equation}
 \begin{figure}
\includegraphics[width=14cm]{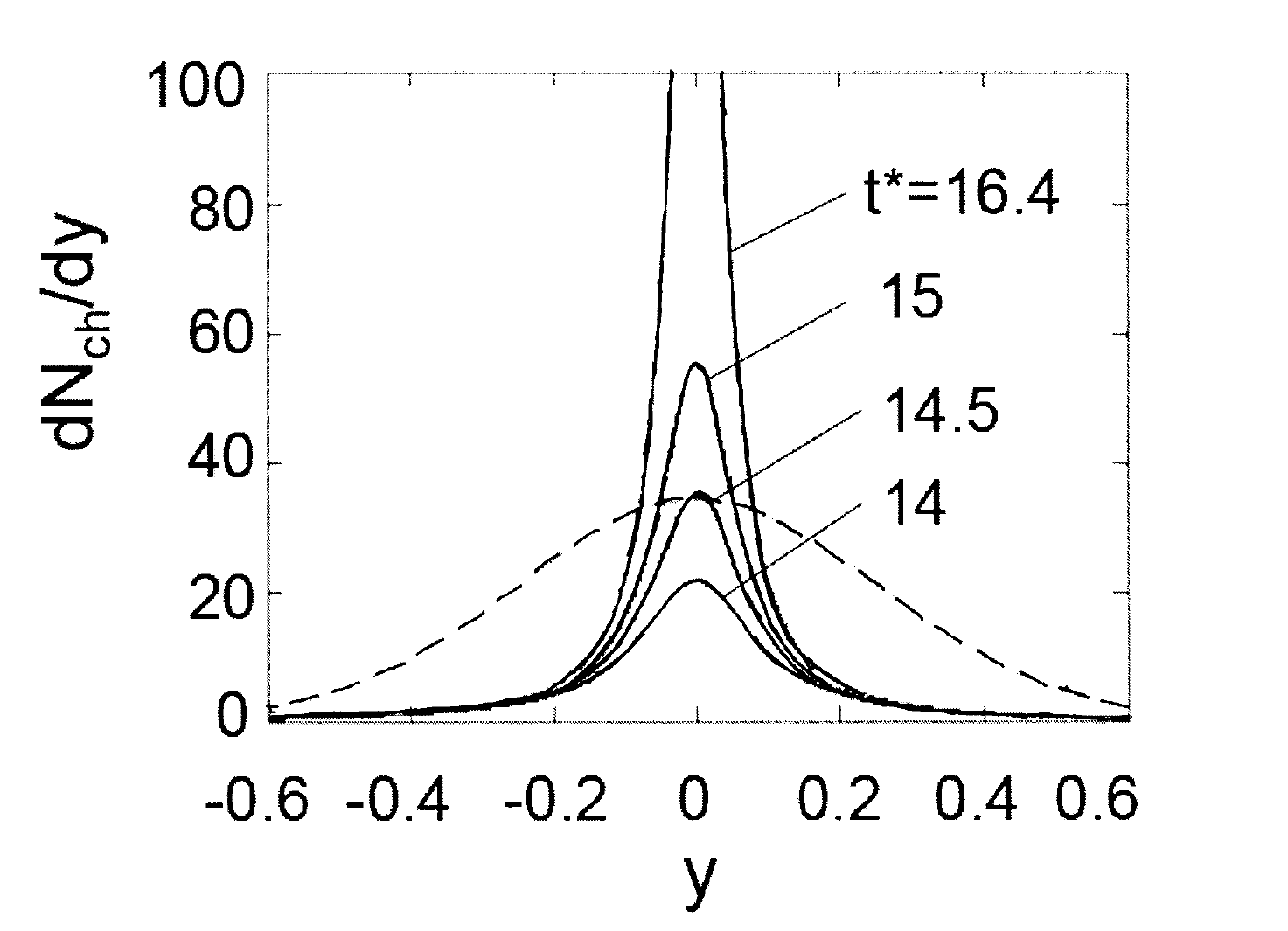}
\caption{Time evolution of rapidity distributions for
produced charged particles from minimum-bias
d + Au collisions at $\sqrt{s_{NN}}$ = 200 GeV in the
initial nonlinear phase for the midrapidity source.
The norm of the distribution is rising with
increasing $t^{*}=D_{y}^{p}\cdot t$ until at $t=\tau_{p}$
all the particles have been produced and the
linear diffusion process in $y-$space starts.
The dashed curve is a Gaussian with the same
particle content $N_{ch}^{3}=22$ as the power-law solution
of the nonlinear problem at $t=\tau_{p}$.}
\label{fig1}
\end{figure}
The corresponding solutions  (\ref{snld}) of the nonlinear diffusion problem
are shown in Fig. 1 for the central source of parton production in d+Au
at $\sqrt{s_{NN}}$= 200 GeV. Results of
the rapidity distribution functions for produced particles multiplied 
with the number of charged hadrons produced in the central source 
$(N_{ch}^{3}=22,$ \cite{wols06})  are shown for four
values of $t^{*}=t\cdot D_{y}^{p}$=14, 14.5, 15, 16.4, with the norm
of P(y,t) reaching $\int P(y,t) dy$=1 for t=$\tau_{p}$=0.25 fm/c at $t^{*}$=16.4.

The distribution functions are seen to rise strongly with increasing $t^{*}$
in a narrow midrapidity region due to the fast increase of the
particle number, with an only moderate increase in the
power-law tails where the particles with higher rapidities are created:
these tails are already present at very short times. Physically, the
partons with the highest rapidity values are created already at the
shortest times.

The rapidity diffusion coefficient in the particle production phase is thus
$D_{y}^{p}=16.4 /\tau_{p}\simeq (16.4/0.25) 
{\mathrm{c/fm}} \simeq  66 {\mathrm{c/fm}}$, 
which is significantly larger than the 
diffusion coefficient $D_{y}$
in the subsequent long-lasting ($\simeq$ 7 fm/c)
linear diffusion phase. This reflects the larger number of degrees of
freedom in the initial phase which is mainly partonic, and the higher
density of particles.

The dashed curve in Fig.1 is a Gaussian that arises from a linear time evolution
($\kappa$=0) and $\delta-$function initial conditions with instant
particle production at $t=0$. It has the same
particle content (22 charged hadrons) as the nonlinear solution
at $t^{*}=\tau_{p}D_{y}^{p}=16.4$. 

For the beam-like sources at initial rapidities $y_{1,2}=\mp y_{max}$, the solutions of
the nonlinear problem for $\kappa=-1$ are accordingly 
($t^{*}=t\cdot D_{y1,2}^{p}\rightarrow t$)
\begin{equation}
P_{1,2}(y,t)=[C_{2}t+(y\pm y_{max})^{2}/(2t)]^{-1}. 
\label{snld12}
\end{equation}
Here the diffusion coefficients in the particle production phase
$D_ {y 1,2}^{p}$ are likely to differ from $D_{y}^{p}$ in the 
midrapidity region because the microscopic processes during
particle production are substantially different, with the diffusion
coefficient at midrapidity mostly due to gluon-gluon collisions.
\begin{figure}
\includegraphics[width=14cm]{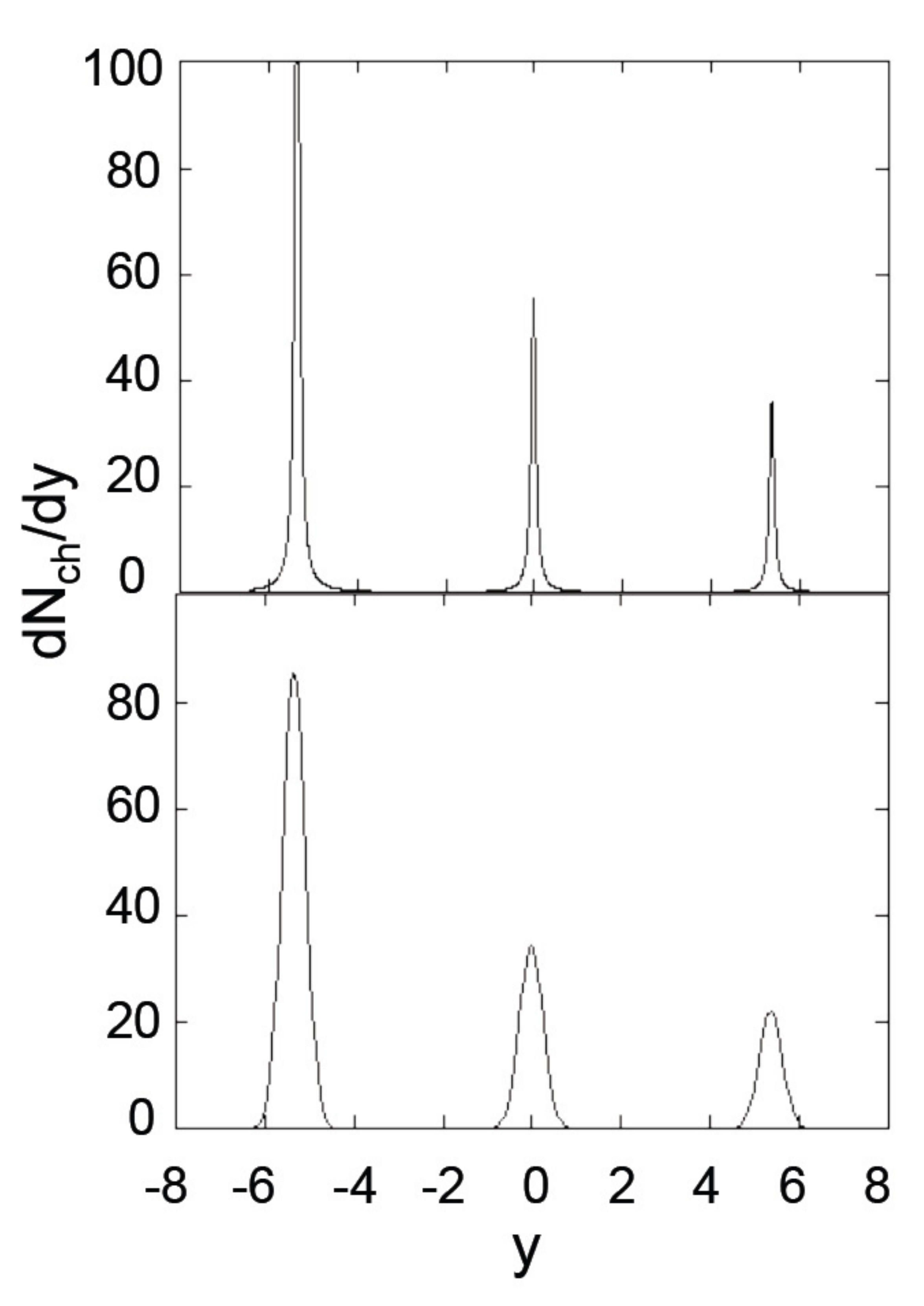}
\caption{Rapidity distributions for
produced charged particles from minimum-bias
d + Au collisions at $\sqrt{s_{NN}}$ = 200 GeV at the end of the
initial nonlinear phase (top) for the three sources, with particle
contents $N_{ch}^{1}=55, N_{ch}^{2}=14,$ and $N_{ch}^{3}=22$.
In the bottom frame, corresponding Gaussians 
are displayed (see also Fig.\ref{fig3}). They arise from a linear evolution with
$\delta-$function initial conditions.}
\label{fig2}
\end{figure}
The analytical solutions at the end of the initial phase $t=\tau_{p}$ in the three
sources are displayed in the upper part of Fig. 2 for minimum-bias d+Au at
$\sqrt{s_{NN}}$= 200 GeV. 

The corresponding number 
of charged hadrons \cite{wols06}
created in the Au-like source is $N_{ch}^{1}$=55, in the d-like source
$N_{ch}^{2}$=14, and in the central source  $N_{ch}^{3}$=22, with a total of
91 produced charged hadrons in minimum-bias collisions.
These particle numbers have been determined from a detailed
comparison of the subsequent linear diffusion phase with
data, see Section \ref{lin}.  
In the lower  part of Fig. 2, Gaussians with the same 
particle-number content are displayed,
as they arise from a linear diffusive time evolution ($\kappa=0$) with
$\delta-$function initial conditions, see next section. 
 
At $t=\tau_{p}$ the power-law solution of the initial nonlinear phase
in the midrapidity source can be expressed as 
\begin{equation}
P_{3}(y,t)=\frac{C}{a^{2}+y^{2}} 
\label{snld3a}
\end{equation}
with
\begin{equation}
C=2\tau_{p}D_{y}^{p}
\label{C}
\end{equation}
\begin{equation}
a=\tau_{p}D_{y}^{p}\sqrt{2\exp(-\tau_{p}D_{y}^{p})}
\label{a}
\end{equation} 
for the central source, and analogously $P_{1,2}(y,t)$ for the beam-like sources with
$y \rightarrow y\pm y_{max}$ and $D_{y}^{p} \rightarrow D_{y1,2}^{p}$. 

As was shown in \cite{wol99,biy02,wols06,wols07,kw07}, the linear diffusive evolution
after the initial short parton production phase is in very good agreement
with the available data. Hence the power-law result (\ref{snld3a}) of the
first phase can be used as an initial condition for the second, 
essentially linear phase that is reconsidered
as described by Eq. (\ref{fpe}) in the next section. With this initial condition, the
solution of the linear diffusion problem for the
central source becomes  \cite{mt08}  
\begin{equation}
R(y,t)=\frac{C}{2a}\sqrt{\frac{2\pi}{\sigma_{y}^{2}}}\Re\Bigl[\exp{\Bigl[-\frac{(y+iv)^{2}}{2\sigma_{y}^{2}}}
\Bigr]\mathrm{erfc}(\frac{iy-v}{\sqrt{2\sigma_{y}^{2}}})\Bigr]
\label{rini}
\end{equation}
with the variance
\begin{equation}
\sigma_{y}^{2}(t)=D_{y}^{p}\tau_{y}[1-\exp(-2t/\tau_{y})]
\label{var}
\end{equation}
and
\begin{equation}
v=a\exp(-t/\tau_{y}).
\label{v}
\end{equation}
For $t\rightarrow \infty$ we have
\begin{equation}
\Re\Bigl[\mathrm{erfc}(\frac{iy}{\sqrt{2\sigma_{y}^{2}}})\Bigr]=1
\label{erf}
\end{equation} 
such that the gaussian limit is attained
\begin{equation}
R(y,t\rightarrow\infty)=\sqrt{\frac{1}{2\pi\sigma_{y}^{2}}}
\exp{\Bigl[-\frac{y^{2}}{2\sigma_{y}^{2}}\Bigr]}
\label{gau}
\end{equation} 
which is a solution of a linear Fokker-Planck equation (\ref{fpe})
for $t\rightarrow \infty$ 
with $\delta-$function initial conditions.
For gaussian initial conditions with an initial variance $\sigma_{0}^{2}$,
one obtains a gaussian solution with $\sigma_{y}^{2}(t)\rightarrow
\sigma_{y}^{2}(t)+\sigma_{0}^{2}exp(-2t/\tau_{y})$
and hence, Eq.(\ref{gau}) results for $t\rightarrow \infty$ as well.

To connect the diffusion approach with data, the linear Relativistic
Diffusion Model (RDM) for the second, long-lasting diffusion phase in 
pseudorapidity space \cite{wol99,biy02,wols06,wols07,kw07}
is reviewed in the next section. 
\section{Linear diffusion phase}
\label{lin}
  
Since the initial power-law behaviour is superseded by the Gaussian evolution
at sufficiently large times, the evolution is started here 
at t=0 with $\delta-$function or gaussian initial conditions to illustrate the outcome
of the three-sources model for large times and in particular, to
compare to data. 

The situation at moderate times with $\delta-$function initial conditions is 
shown in the lower part of Fig. 2, with separate Gaussians in the 
three sources in rapidity space which have the same particle-number
content as the initial power-law solutions of the nonlinear
particle-production problem. The subsequent time evolution
of these three sources leads to agreement with the
experimental data at the interaction time $t=\tau_{int}$,
and to statistical equilibrium for $t\rightarrow \infty$.
  
The nonequilibrium-statistical description of this evolution
is based on an essentially linear diffusion equation which
is briefly reviewed in this section.
We have used a Fokker-Planck
equation (Uhlenbeck-Ornstein \cite{uo30} version with $\kappa=0$)
\cite{wol99,biy02,wols06,wols07,kw07} for the 
distribution function $R(y,t)$ for produced charged hadrons
in rapidity space
\begin{equation}
\frac{\partial}{\partial t}R(y,t)=-\nabla_{y}\Bigl[J(y)R(y,t)\Bigr]+
D_{y}\nabla_{y}^{2}R(y,t).
\label{lfpe}
\end{equation}
The drift is now taken into account since we look at the
large-time behaviour, and the
drift function $J(y)$ determines the speed of the 
statistical equilibration in $y$-space.
In order to attain the Boltzmann distribution for large times,
the drift term must have the form \cite{alb00,wol03}       
\begin{equation}
J(y)\propto m_{\perp}\sinh(y)\propto p_{\parallel}
\label{J}
\end{equation}
with the transverse mass $m_{\perp}=\sqrt(m^{2}+p_{\perp}^{2})$,
and the longitudinal momentum $p_{\parallel}$. This introduces another
nonlinearity into the problem, which prohibits an analytical solution of the
diffusion equation. Such an analytical solution \cite{wol99} is, however, possible
by linearising the drift function in a relaxation ansatz
\begin{equation}
J(y)=(y_{eq}-y)/\tau_{y}
\label{J}
\end{equation}
with the rapidity relaxation time $\tau_{y}$, and the equilibrium value
of the rapidity $y_{eq}$
that is calculated from energy and momentum conservation
in the system of participants. The deviations of the solutions for nonlinear
and linear versions of the drift are not pronounced \cite{wol03} and hence,
I have used the analytical solutions of the linear problem for the
components $R_{k}(y,t)$ ($k$=1,2,3) of the distribution function
\begin{equation}
\frac{\partial}{\partial t}R_{k}(y,t)=
\frac{1}{\tau_{y}}\frac{\partial}
{\partial y}\Bigl[(y-y_{eq})\cdot R_{k}(y,t)\Bigr]
+D_{y}^{k}\frac{\partial^2}{\partial y^2}
R_{k}(y,t).
\label{fpe}
\end{equation}
The diagonal components $D_{y}^{k}$ of the rapidity
diffusion tensor contain the
microscopic physics in the respective beam-like $(k=1,2)$
and central $(k=3)$ regions. They account for the statistical broadening of the
distribution functions. To connect with data, one has to consider the
additional broadening due to longitudinal collective expansion that leads
to a larger (effective) value of $D_{y}$ \cite{wol06} than what is 
calculated \cite{wol99} from
the dissipation-fluctuation theorem (Einstein relation).

As discussed  above, the initial conditions in the linear phase are taken as
$R_{1,2}(y,t=0)=\delta(y\pm y_{max})$
with the maximum rapidity $y_{max}=5.36$ at
the highest RHIC energy of $\sqrt{s_{NN}}$ = 200 GeV
(beam rapidities are $y_{1,2}=\mp y_{max}$), and $R_{3}(y,t=0)=\delta(y).$
A midrapidity gluon-dominated symmetric source had also been proposed by
Bialas and Czyz \cite{bc05}. 

This initial condition for the midrapidity source in the linear case corresponds to 
initial particle production without any longitudinal motion, independently of the mass
of the collision partners: the third source is created at $y=0$, and
the drift towards the equilibrium
value $y=y_{eq}$, as well as the rapid collective expansion,
sets in subsequently. (In contrast, the nonlinear model as
discussed in the previous section produces power-law tails
at short times, which are superseded by the gaussian tails
of the linear evolution only at later times).

The mean values in the three sources have the time dependence
\begin{equation}
<y_{1,2}(t)>=y_{eq}[1-\exp(-t/\tau_{y})] \mp y_{max}\exp{(-t/\tau_{y})}
\label{mean}
\end{equation}
for the sources (1) and (2), and 
\begin{equation}
<y_{3}(t)>=y_{eq}[1-\exp(-t/\tau_{y})]
\label{mean3}
\end{equation}
for the moving central source. The three mean values reach the equilibrium
limit for time to infinity. In our previous RDM-calculation \cite{wols06}
with slightly different initial condition, 
the mean value of the central source was at the equilibrium limit 
$<y_{3}(t)>=y_{eq}$ independently of time, thus assuming instant
equilibration in this source regarding the mean values.
The variances $\sigma_{1,2,3}^{2}(t)$ are as in Eq.(\ref{var}),
with $D_{y}\rightarrow D_{y}^{p}$.

It  turns out that for d+Au at the highest RHIC energy, one can not
determine from a comparison with the data which of the two possibilities
for the initial conditions of the central source is more realistic because
the $\chi^{2}$ is nearly identical in both cases.
The subsequent diffusion-model time evolution
in pseudorapidity space is followed up to the interaction time $\tau_{int}$,
when the produced charged hadrons cease to interact
strongly. 

The quotient $\tau_{int}/\tau_{y}$ is determined from the minimum $\chi^{2}$
with respect to the data, simultaneously with the minimization
of the other free parameters - namely, the
variances of the three partial distribution functions,
and the number of particles produced in the central source.
In this nonequilibrium-statistical approach,
the equilibrium value of the rapidity and its dependence on
centrality is calculated from energy and momentum conservation
in the system of participants as
\begin{eqnarray}
y_{eq}(b)=
\frac{1}{2}\ln\frac{<m_{1}^{T}(b)>\exp(-y_{max})+<m_{2}^{T}(b)>
\exp(y_{max})}
{<m_{2}^{T}(b)>\exp(-y_{max})+<m_{1}^{T}(b)>\exp(y_{max})}
\label{yeq}
\end{eqnarray}\\
with
the transverse masses $<m_{1,2}^{T}(b)>=
\sqrt(m_{1,2}^2(b)+
<p_{T}>^2)$, and masses
m$_{1,2}(b)$ of the "target"(Au)- and "projectile"(d)-participants 
that depend on the impact parameter $b$. The average 
numbers of participants 
$<N_{1,2}(b)>$ from the Glauber
calculations reported in \cite{bbb04} for minimum bias d + Au 
at the highest RHIC energy are
 $<N_{1}>=$6.6, $<N_{2}>=$1.7,
which we had also used in \cite{wols06}.

The average numbers of charged particles in
the target- and projectile-like regions $N_{ch}^{1,2}$ are 
proportional to the respective
numbers of participants $N_{1,2}$,
\begin{equation}
N_{ch}^{1,2}=N_{1,2}\frac{(N_{ch}^{tot}-N_{ch}^{eq})}{(N_{1}+N_{2})}
\label{nch}
\end{equation}
with the constraint $N_{ch}^{tot}$ = $N_{ch}^1$ + $N_{ch}^{2}$ +
$N_{ch}^{eq}$.
Here the total number of charged particles 
$N_{ch}^{tot}$ is determined from the data. The average number
of charged particles in the equilibrium source $N_{ch}^{eq}$ is a
free parameter that is optimized together with the variances
and $\tau_{int}/\tau_{y}$ in a $\chi^{2}$-fit of the data
using the CERN minuit-code. The results are summarized in table 1.
\begin{table}
\caption{Produced charged hadrons 
 in minimum-bias d + Au  
collisions at $\sqrt{s_{NN}}$ =
200 GeV, y$_{1,2}=\mp$ 5.36  in the linear Relativistic Diffusion Model.
The equilibrium value of the rapidity in the RDM
is $y_{eq}$, the time parameter (see text) is $p$,
the corresponding value of interaction time over relaxation
time is $\tau_{int}/\tau_{y}$, the variance of the central source in
$y-$space is $\sigma_{3}^{2}$. The number of produced charged particles 
is $N_{ch}^{1,2}$ for the sources 1 and 2 and $N_{ch}^{3}$ for the central
source, the percentage of
charged particles produced in the midrapidity source is $n_{ch}^{3}$,
and $\chi^{2}/d.o.f.$ is the result of the minimization \cite{wols06} per
number of degrees of freedom.}
\vspace{.1cm}
\label{tab1}
\begin{tabular}{ccccccccc}
\hline
$y_{eq}$&$p$&$\tau_{int}/\tau_{y}$&
$\sigma_{3}^{2}$&$N_{ch}^{1}$&$N_{ch}^{2}$&$N_{ch}^{3}$&$n_{ch}^{3}$(\%)&$\chi^{2}/d.o.f.$\\
\hline
- 0.664&0.54&0.78&4.19 &  55 & 14& 22 & 24&2.44/48 \cr 
\hline
\end{tabular}
\end{table}

The FPE is solved analytically as outlined in \cite{wol99},
and the solutions are converted to pseudorapidity space. 
This conversion is required because
particle identification is not available. The relation between scattering
angle $\theta$ and pseudorapidity $\eta$ is
$\eta=-$ln[tan($\theta / 2)]$. Here $\theta$ is
measured relative to the direction of the deuteron beam. Hence,
particles that move in the direction of the gold beam have negative, particles 
that move in the deuteron direction have positive
pseudorapidities.
The conversion from $y-$ to $\eta-$
space of the rapidity density
\begin{equation}
\frac{dN}{d\eta}=\frac{p}{E}\frac{dN}{dy}=
j(\eta,\langle m\rangle/\langle p_{T}\rangle)\frac{dN}{dy} 
\label{deta}
\end{equation}
is performed through the Jacobian
\begin{equation}
j(\eta,\langle m\rangle/\langle p_{T}\rangle) = \cosh({\eta})\cdot 
[1+(\langle m\rangle/\langle p_{T}\rangle)^{2}
+\sinh^{2}(\eta)]^{-1/2}.
\label{jac}
\end{equation}The average mass 
 $<m>$ of produced charged hadrons in the
central region is approximated by the pion mass $m_{\pi}$, and a
mean transverse momentum $<p_{T}>$ = 0.4
GeV/c is used \cite{wols06}. Due to the conversion,
the partial distribution functions are different from Gaussians.
The charged-particle distribution in rapidity space is obtained
as incoherent 
superposition of nonequilibrium and local equilibrium solutions of
(\ref{fpe}) 
\begin{equation}
\frac{dN_{ch}(y,t=\tau_{int})}{dy}=N_{ch}^{1}R_{1}(y,\tau_{int})
+N_{ch}^{2}R_{2}(y,\tau_{int})
+N_{ch}^{3}R_{3}(y,\tau_{int})
\label{normloc1}
\end{equation}
with the interaction time $\tau_{int}$ (total integration time of the
differential equation). The integration is 
stopped at the value of $\tau_{int}/\tau_{y}$ that produces the
minimum $\chi^{2}$ with respect to the data and hence, the
explicit value of $\tau_{int}$ is not needed as an input. 
The resulting values for $\tau_{int}/\tau_{y}$ are given
in table \ref{tab1} together with the widths of the central
 distributions, and the particle numbers in the three sources.
  \begin{figure}
\includegraphics[height=19.5cm]{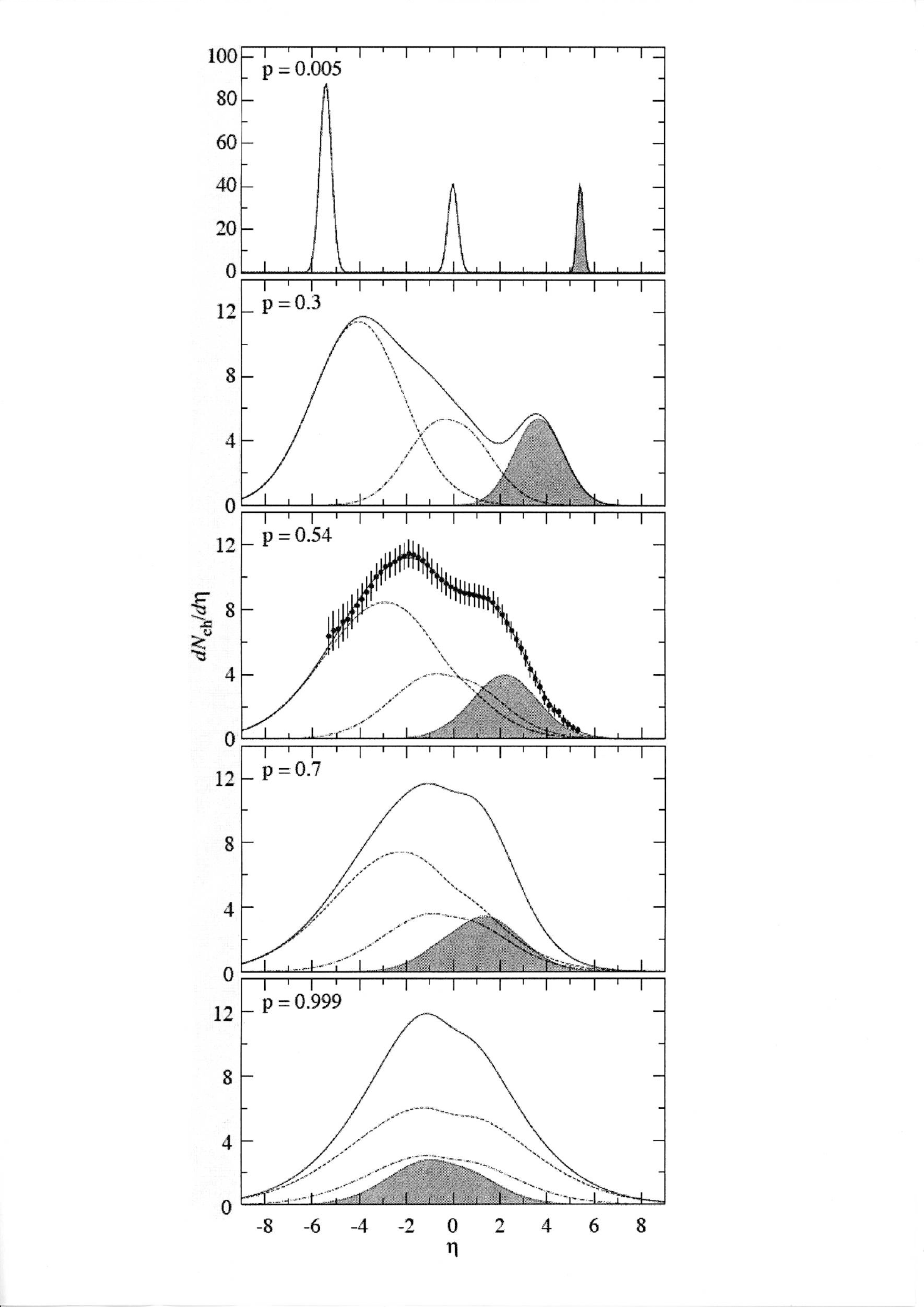}
\caption{Time evolution of pseudorapidity distributions for
produced charged particles from minimum-bias
d + Au collisions at $\sqrt{s_{NN}}$ = 200 GeV in the
linear diffusion model as outlined in \cite{wols06,wols07}.
The d-like source is shaded to illustrate the
movement in $\eta-$space towards equilibrium.
Dash-dotted curves show the slightly moving, gluon-dominated
midrapidity source for hadron production.
Results for five
time steps ($p$-values, cf. text) are displayed.
Agreement with the data \cite{bbb04} is reached at $p$=0.54.
Statistical equilibrium centered at $\eta_{eq}$
would be achieved at later times. Reviewed from \cite{wols07}.} 
\label{fig3}
\end{figure}
 
The time evolution is shown together with the comparison to 
 PHOBOS minimum-bias data \cite{bbb04} in 
 Fig.\ref{fig3}. It is evident that the two beam-like distribution
 functions move towards smaller absolute pseudorapidities as time
 increases, and reach agreement with the data at $p$=0.54.
 Here the time evolution parameter $p$
in the numerical calculation is defined as
\footnote{
There is a difference of 
a factor of two in the exponent as compared to the definition of $p$ used in 
\cite{wols06}, which causes different $t/\tau_{y}$ values for given $p$.}
\begin{equation}
p=1-\exp(-t/\tau_{y}) .
\label{pe}
\end{equation}

The minimum-bias result also
shows the asymmetric shape of the distribution function, which is very well reproduced
in the diffusion calculation. At larger values of the time evolution
parameter $p$, all three subdistributions tend to become symmetric
in $y$ with respect to the equilibrium value $y_{eq}$, indicating the approach
to thermal equilibrium. At $p$=0.999, the equilibrium state is already
closely approached. The slight asymmetry is due to the conversion from
rapidity- to pseudorapidity space which tends to produce a dip at
$\eta=0$. For time to infinity,
statistical equilibrium in pseudorapidity space would be reached.

\begin{figure}
\includegraphics[height=19.5cm]{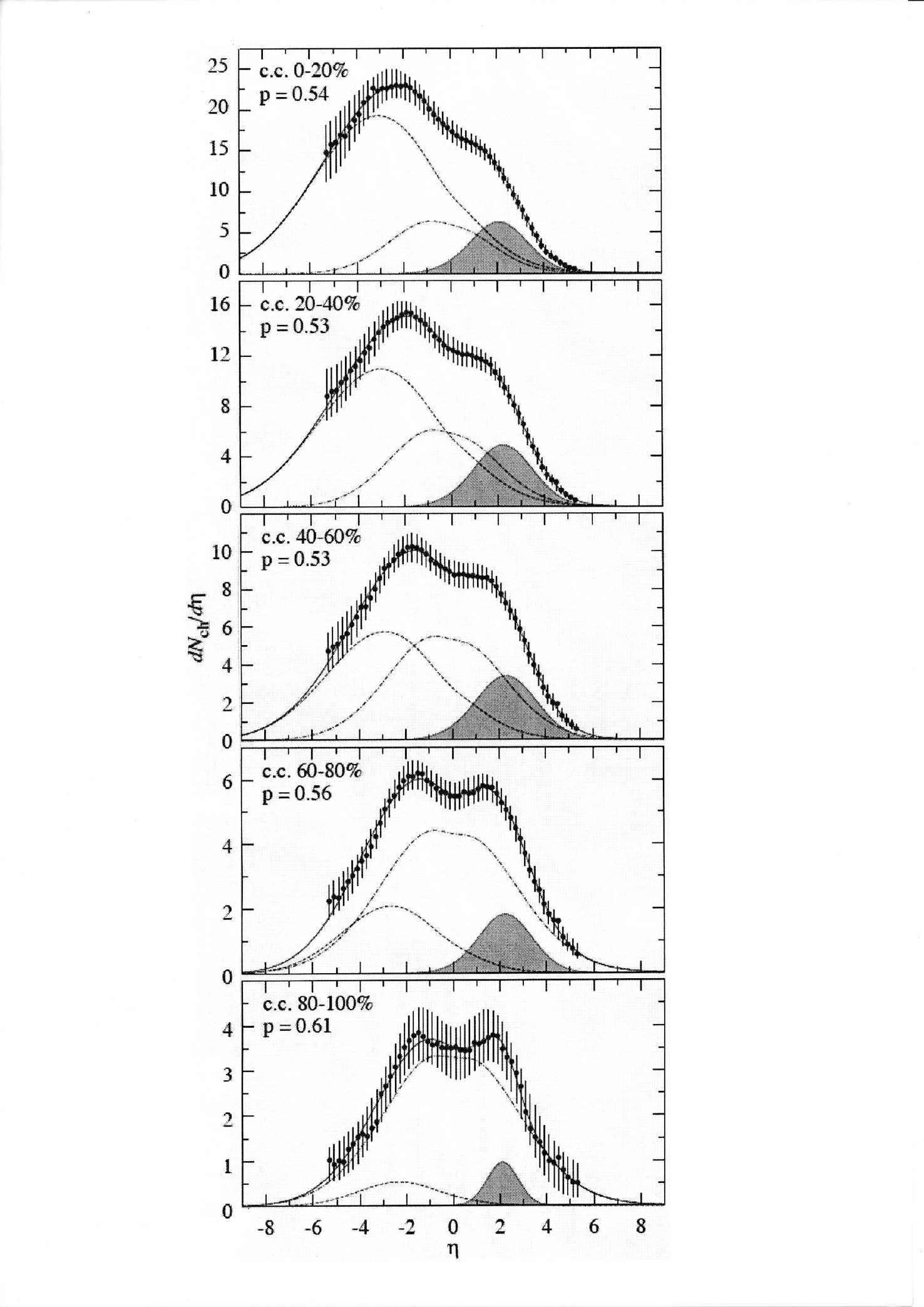}
\caption{Calculated pseudorapidity distributions of 
produced charged particles in
 d + Au collisions at $\sqrt{s_{NN}}$ = 200 GeV for five
different centralities,
as outlined in \cite{wols06,wols07}.
Central collisions are shown in the top
frame, peripheral at the bottom.
The linear diffusion-model (RDM) results for three sources
(d-like source shaded) in $\eta-$space
are shown together with their incoherent sums as $\chi^{2}-$
minimizations at each centrality cut (c.c.).
The time variable is $p$ (see text).
The initial conditions for the central source are slightly different
from \cite{wols06}, see text.
Data are from PHOBOS \cite{bbb04}. Reviewed from \cite{wols07}.} 
\label{fig4}
\end{figure}

We have shown in \cite{wols06} that with this linear RDM approach, the 
centrality dependence of the measured pseudorapidity 
distributions \cite{bbb04} from
central to very peripheral collisions can also be modeled in
considerable detail, Fig.\ref{fig4}. For peripheral collisions, the asymmetry
of the overall distribution is not yet pronounced because here the
d- and the Au-like partial distributions are similar in size due
to the small number of participants. 

Towards more central collisions,
the number of gold participants rises, and the corresponding
partial distribution of produced particles becomes more important.
In addition, the distributions drift towards the equilibrium value.
Both effects produce the asymmetric shape, which is also 
seen in minimum-bias. The tails of the distribution functions
are of gaussian shape in perfect agreement with the data.
This shows that the power-law tails of the initial phase have not survived the
time evolution, as is confirmed when the result of the initial phase
is used explicitly as an initial condition for the linear evolution.

It is interesting to compare the behaviour of the rapidity
or pseudorapidity distribution functions 
with results from different
approaches to the problem such as saturation models 
\cite{arm05,nar05,mcl94},
viscous hydrodynamics \cite{den07}, or ideal hydrodynamics \cite{cso03}.

Calculations within the framework of 
the Parton Saturation Model not only predict the midrapidity 
value, but also the full rapidity distribution function (at RHIC energies, and
also at LHC) \cite{nar05,kha05}. These calculations are based on a classical 
effective theory that describes the gluon distribution in large nuclei 
at high energies where saturation might occur at a critical momentum 
scale, to form a Color Glass Condensate (CGC) \cite{mcl94}.

This assumption has a clear and reasonable physical 
basis and yields good results for pseudorapidity 
distribution functions of produced charged hadrons 
at the available RHIC energies. Problems may be expected
for net-proton rapidity distributions since protons
and antiprotons are produced in equal amounts from the CGC.
At LHC
energies, the overall pseudorapidity distribution from the CGC
as obtained with the assumption of a constant $\alpha_{s}$ for strong 
coupling is slightly narrower than the corresponding diffusion-model 
prediction \cite{kw07}.

Additional consideration of a running coupling gives a midrapidity value that
is of the order of 10{\%} smaller; another uncertainty arises from the 
extrapolation of the saturation scale to LHC 
energies. Various predictions for central rapidity densities and
pseudorapidity distributions at RHIC and LHC energies had been
summarized e.g. in \cite{arm00}, where also the differences among the existing
models - including hydrodynamical and pQCD approaches and 
their numerical implementations - had been discussed. 

In relation to these approaches,
the analytical diffusion model provides good results
when compared in detail to the experimental distribution functions
at RHIC energies, in particular, in the tails of the distributions.
To provide a microscopic
foundation, however, a derivation of the diffusion coefficients in
the three sources would be required. Due to the valence-quark dominance
in the beam-like sources as opposed to the mainly gluonic
midrapidity source, the diffusion coefficients may turn
out to be substantially different in the three sources.

\section{Conclusion}
\label{conc}
I have presented a nonlinear Relativistic Diffusion Model that
includes an explicit analytical treatment of the initial parton-production
phase in rapidity space. As a consequence of the high rapidity density at
short times $t \simeq 0.25$ fm/c, the rapidity diffusion
coefficient depends on the distribution function, such that the 
problem is highly nonlinear in the initial phase. 

For a power-law dependence of the diffusion coefficient
on the distribution function with an exponent $\kappa$,
the mathematical technique of similarity
solutions that has been refined in many previous works
proves to be useful in the present physical context. 
In particular, I have investigated the so-called source 
solution in rapidity space. 

An adequate solution to the nonlinear problem
should not have a free boundary in rapidity
space that moves with finite velocity as is the case for positive 
values of $\kappa$, but in parton production there should be
an instantaneous spread in $y-$space without a free boundary.
This corresponds to the case $\kappa=-1$, which can be
solved analytically using the technique of similarity solutions.

During particle production in three sources, the norm of the distribution function
increases, which I have considered phenomenologically by
letting the integration constant depend on particle number.
Since particle production is very rapid - exponentially in time -,
this increase of the norm of the distribution function turns out to be faster
than the spread of the distribution function in rapidity space
due to nonlinear diffusion and hence, the power-law tails of the
distribution function remain small during the parton-production phase.

The  result of the parton-production phase is then used
as initial condition for the later (pre-hadronic and hadronic)
stage of the collision, which is treated here in the linear
relativistic diffusion model. The linear evolution washes out the
initial power-law tails of the distribution function, which
become gaussian. 

With the proper Jacobi  transformation
to pseudorapidity space, this approach yields very precise
agreement with charged-hadron data for both asymmetric 
systems (d+Au), and symmetric
systems such as Cu+Cu and Au+Au \cite{biy02,kw07}.
It is also particularly suitable for predictions at LHC
energies of 5.5 TeV for Pb+Pb.

The second collision phase lasts about
7-10 fm/c depending on the system, the incident energy, and
the centrality. Due to the schematic treatment
that is based on a linear partial differential equation, particle
number is conserved in this phase, which appears as a reasonable
physical assumption even though it is certainly not strictly valid.

For time to infinity, the evolution of the distribution function 
proceeds to statistical equilibrium with respect to the variable
rapidity or pseudorapidity. Comparing the data with this
time evolution as modeled by the solutions of the linear problem,
it is evident that at the time of the measurement - when strong interaction
ceases - the system is still far from statistical equilibrium.
This underlines the view that relativistic heavy-ion collisions
are very suitable to observe strongly interacting many-body systems
with a large amount of particle production
on their way to statistical equilbrium. 

In this work I have not considered the connection between the 
diffusion approach and QCD. It is obvious that the forward and backward 
sources for produced particles are related to the valence quarks, whereas the
central source is essentially due to gluon-gluon collisions. An actual
microscopic calculation of the three sources emphasizing their relative size
(number of produced particles) is therefore of interest \cite{08}.

\bf{Acknowledgements}

\rm
I thank
M. Biyajima, T. Mizoguchi and N. Suzuki for our
collaboration within the linear Relativistic Diffusion Model (RDM)
as presented in \cite{wols06,wols07} and reviewed here in Section 4, 
and Y. Mehtar-Tani for discussions
and for the derivation of Eq. (\ref{rini}). The project is supported by DFG 
under contract No. STA 509/1-1.



\end{document}